\documentclass[11pt]{article}
\usepackage[a4paper]{geometry}
\usepackage[english]{babel}

\usepackage{amsmath, accents}
\usepackage{amsfonts}
\usepackage{amstext}
\usepackage{amssymb}
\usepackage{amsthm}
\usepackage{amscd}
\usepackage{slashed}

\usepackage[pagebackref,draft=false]{hyperref}
\hypersetup{colorlinks,
linkcolor=myrefcolor,
citecolor=mycitecolor,
urlcolor=myurlcolor}

\usepackage[capitalize]{cleveref}
\usepackage{cite}
\usepackage{caption}
\usepackage{etaremune}

\usepackage{xcolor}
\definecolor{myurlcolor}{rgb}{0,0,0.4}
\definecolor{mycitecolor}{rgb}{0,0.5,0}
\definecolor{myrefcolor}{rgb}{0.5,0,0}
\usepackage{graphicx}
\usepackage{tikz}
\usepackage{tikz-cd}
 \usetikzlibrary{decorations.pathmorphing}

\usepackage{mathrsfs}
\newcommand{\be}{\begin{equation}}
\newcommand{\ee}{\end{equation}}
\newcommand{\bea}{\begin{eqnarray}}
\newcommand{\eea}{\end{eqnarray}}

\newcommand{\ra}{\rightarrow}

\newcommand{\EL}{\mathscr{E}\hspace{-0.05cm}\mathscr{L}}

\newcommand{\dd}{{\rm d}}

\title{Covariant reduction of classical Hamiltonian Field Theories: \\ From D'Alembert to Klein-Gordon and Schr\"odinger}
\date{}

\author{F. M. Ciaglia$^{1,7}$ \href{https://orcid.org/0000-0002-8987-1181}{\includegraphics[scale=0.7]{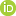}}, F. Di Cosmo$^{2,3,8}$ \href{https://orcid.org/0000-0003-0256-5913}{\includegraphics[scale=0.7]{ORCID.png}}, A. Ibort$^{2,3,9}$ \href{https://orcid.org/0000-0002-0580-5858}{\includegraphics[scale=0.7]{ORCID.png}}, \\ G. Marmo$^{4,5,10}$ \href{https://orcid.org/0000-0003-2662-2193}{\includegraphics[scale=0.7]{ORCID.png}}, L. Schiavone$^{3,4,6,11}$  \href{https://orcid.org/0000-0002-1817-5752}{\includegraphics[scale=0.7]{ORCID.png}} \\
\footnotesize{$^{1}$\textit{ Max Planck Institute for Mathematics in the Sciences, Leipzig, Germany}} \\
\footnotesize{$^{2}$\textit{ ICMAT, Instituto de Ciencias Matem\'{a}ticas (CSIC-UAM-UC3M-UCM)}} \\
\footnotesize{$^{3}$\textit{Depto. de Matem\'aticas, Univ. Carlos III de Madrid, Legan\'es, Madrid, Spain}} \\
\footnotesize{$^{4}$\textit{ INFN-Sezione di Napoli, Naples, Italy}} \\
\footnotesize{$^{5}$\textit{ Dipartimento di Fisica ``E. Pancini'', Universit\`a di Napoli Federico II,  Naples, Italy}} \\
\footnotesize{$^{6}$\textit{ Dipartimento di Matematica e Applicazioni "Renato Caccioppoli", Università di Napoli Federico II, Napoli, Italy}} \\
\footnotesize{$^{7}$\textit{ e-mail: \texttt{florio.m.ciaglia[at]gmail.com} and \texttt{ciaglia[at]mis.mpg.de}}} \\
\footnotesize{$^{8}$\textit{ e-mail: \texttt{fcosmo[at]math.uc3m.es}}} \\
\footnotesize{$^{9}$\textit{ e-mail: \texttt{albertoi[at]math.uc3m.es}}} \\
\footnotesize{$^{10}$\textit{ e-mail: \texttt{marmo[at]na.infn.it}}} \\ 
\footnotesize{$^{11}$\textit{ e-mail: \texttt{luca.schiavone[at]unina.it}}}  
}

\begin{document}

\maketitle

\begin{abstract}
A novel reduction procedure for covariant classical field theories, reflecting the generalized symplectic reduction theory of Hamiltonian systems, is presented.   
The departure point of this reduction procedure consists in the choice of a submanifold of the manifold of solutions of the equations describing a field theory.
Then, the covariance of the geometrical objects involved, will allow to define equations of motion on a reduced space. 
The computation of the canonical geometrical structure is performed neatly by using the geometrical framework provided by the multisymplectic description of covariant field theories.  
The procedure is illustrated by reducing the D'Alembert theory on a five-dimensional Minkowski space-time to a massive Klein-Gordon theory in four dimensions and, more interestingly, to the Schr\"odinger equation in 3+1 dimensions.
\end{abstract}

\section{Introduction}

A novel covariant reduction procedure of classical Hamiltonian field theories based on the geometry of the space of solutions of the theory is presented.  
Since the seminal contributions of Peierls\cite{Peierls1952-Commutation_relativistic}, de Witt\cite{DeWitt1965-Groups_Fields}, Crnkovic'-Witten\cite{Cr87,Cr88}  (just to cite a few key contributions on the subject)  to the more recent geometrical rendering by Forger-Romero\cite{ForgRomero2005-Covariant_poisson_brackets} and the contributions by Asorey \textit{et al}\cite{A-C-DC-I-2017,A-C-DC-I-M-2017} and Ciaglia \textit{et al}\cite{C-DC-I-M-S-2020-01, C-DC-I-M-S-2020-02}, it has become evident that the natural geometrical structures carried by the space of solutions of the Euler-Lagrange equations of the theory provide   invaluable information of the theory itself both on its quantum aspects (as the classical solutions can be understood as the classical counterpart of the ground state of the quantum theory) and on its symmetries.   

Specifically, it is well-known that the manifold of solutions of the equations describing the dynamics of some Hamiltonian classical field theory, denoted in what follows by $\EL$ and called often ``the covariant phase space'' in the previous literature, carries a canonical presymplectic structure $\Omega$ (see, for instance Ref.\cite{ForgRomero2005-Covariant_poisson_brackets} and, more recently, Ref.\cite{Ci20}).   
Due to the covariance of the tensorial objects involved in this physical descriptions (Homiltonian functionals, differential forms), we may define a generalized reduction procedure which is associated with the choice of submanifolds within $\EL$. 
Indeed, the characteristic distribution of the pullback of the canonical presymplectic structure $\Omega$ to a submanifold $\mathcal{W}\subset \EL$ can be used to define an equivalence relation on $\mathcal{W}$ and one can obtain a reduced dynamics on the quotient space determined by this equivalence relation (for a detailed exposition of reduction procedures see for instance \cite[Chap. 7]{Ca15} and references therein). 

The choice of the submanifold depends on the problem we are analyzing: It may be selected, for instance, either as one of the level sets of constants of the motion or via constraint relations. 
However, it is worthful remarking here that this reduction, involving directly submanifolds of the manifold of solutions $\EL$, has a ``global'' meaning, in the sense that the chosen submanifold needs not to be again the manifold of solutions of the equations describing the dynamics of a ``reduced'' field theory (for instance we are not ensured that any submanifold of the manifold of solutions of the equations of motions of a field theory become the manifold of solutions of the equations of motion for a field theory defined on a lower dimensional spacetime or possessing a reduced space of values).
This analysis can be performed only afterwards and it will depend on the particular physical situation under analysis.

In this letter we will present some examples where the chosen submanifolds of solutions of the equations of motion of a field theory are symplectic submanifolds that can be themselves considered as the manifolds of solutions for the equations of motion of a ``reduced'' field theory. 
In particular, it will be shown that starting from the D'Alembert equation on a five-dimensional Minkowski spacetime, this generalized reduction procedure will lead either to a massive Klein-Gordon equation (KG), or to the Schr\"{o}dinger equation in a space of dimension one unit less.  

This idea has been used, \cite{C-DC-F-M-M-S-V-V-2020} to show how it is possible to obtain nonlinear equations by reducing linear ones, or to obtain approximating solutions of the initial equations. 
When the selected submanifold is an invariant submanifold for the original equations of motion, we may perform also the reduction by restricting directly the equations of motion to the submanifold.
This will be the case of our illustrative examples: Therefore we shall first consider the reduction at the level of evolutionary equations and later we go to their description in covariant terms.
In particular, the covariant description will be performed according to a multisymplectic approach which is suitable for first-order Hamiltonian field theories. 
The geometric features characterizing this formalism, which is based on the definintion of a covariant phase space equipped with a multisymplectic differential form, indeed, are well suited to the reduction procedures involving the choice of a submanifold of the manifold $\EL$. 
As an intermediate step we consider also the reduction of the dispersion relations associated with the principal symbol of the differential operator defining the equations of motion of the initial field theory. 
This procedure, can be thought of as a reduction of the ``Hamiltonian'' description of the particles which are associated to the field.

The idea that the manifolds of solutions of both Klein-Gordon and Schr\"{o}dinger equations can be obtained as reduction of a common unfolded space of solutions is motivated by the fact that both these equations are characterized by their invariance with respect to the Poincar\'{e} and the Galilei group respectively.
Then, a key observation is that these groups are realized as subgroups of the same group $ISO(1,4)$, naturally acting on $\mathbb{R}^{5}$.
This is the main observation which motivates to seek for a five-dimensional equation which restricted to appropriate invariant submanifolds of solutions would give rise to Klein-Gordon or Schr\"{o}dinger equation. 
The sought for equation turns out to be a D'Alembert type equation associated with a Laplace-Beltrami operator in a five dimensional Minkowski spacetime.
Then, the solutions to the original Schr\"{o}dinger equation will be obtained as ``harmonic functions'' which are also eigenfunctions of the infinitesimal generator of the translations along the additional direction which must be light-like.
To recover the solutions for the Klein-Gordon equations, instead, we should restrict to harmonic functions which are also eigenfunctions of the generator of translation along the additional direction which must be space-like.

Before ending this introduction it is worth mentioning that this bundle approach to reduction could be used also for a non-Abelian extension, for instance to introduce isospin or colored charges.
However, additional remarks on this feature will be presented in the conluding section. 

\section{Reduction of evolutionary fields equations}\label{sec: reduction of e.o.m.}

Consider a five-dimensional, globally hyperbolic space-time $(\mathcal{M}, \eta)$, with $\mathcal{M}=\mathbb{R}^{5}$, and with $\eta$  the Lorentzian metric tensor given by 
\be
\eta = \eta_{\mu \nu} \dd x^{\mu} \otimes \dd x^{\nu} = \dd x^0 \otimes \dd x^0 - \delta_{jk}\dd x^j \otimes \dd x^k ,
\ee
where $(x^{\mu})$, $\mu = 0,1, \ldots, 4$, is a global set of Cartesian coordinates on $\mathcal{M}$  .
By means of $\eta$,  we can write the so-called D'Alembert equation
\be
\eta^{\mu\nu}\, \partial_{\mu} \partial_{ \nu}\,\varphi\,=\,0\,
\ee
for a real or complex-valued function $\varphi$ on $\mathcal{M}$.
Let us note that the group $ISO(1,4)$ acts linearly on $\mathcal{M}$ and represents a symmetry group for the D'Alembert equation.
This equation is also invariant under conformal transformations, but these will not concern us here.

If we focus our attention   on those functions  $\varphi$  of the form
\be\label{eqn: KG constraint}
\varphi(x^{0},x^{1},x^{2},x^{3},x^{4})\,=\,\mathrm{e}^{- m x^{4}}\,u(x^{0},x^{1},x^{2},x^{3})\,,
\ee
we immediately obtain that the D'Alembert equation for $\varphi$ reduces to  the Klein-Gordon equation for the function $u$ given by
\be
\partial^{2}_{0}u\,=\,\Delta\,u + m^{2}\,u\,.
\ee
Let us remark  that a subgroup of $ISO(1,4)$ preserving the set of functions selected in  Eq.\eqref{eqn: KG constraint} can be identified with the Poincar\'{e} group.

Now, we can introduce a new set of coordinates $t,s$ on $\mathcal{M}$, called ``advanced'' and ``retarded coordinates'', given by
\be\label{eqn: coordinate cono luce}
t = x^0 - x^4 \, , \qquad s = x^0 + x^4 \, .
\ee
It is easy to check that in the new coordinates the metric becomes
\be\label{eqn: metrica nelle coordinate di cono luce}
\eta = \frac{1}{2}  \dd t\otimes_{S} \dd s -  \sum_{i,j= 1}^ {3}\eta_{ij} (x) \dd x^i \otimes \dd x^j .
\ee
If we focus our attention  only on those functions  $\varphi$ of the form
\be\label{eqn: SE constraint}
\varphi(t,x^{1},x^{2},x^{3},s)\,=\,\mathrm{e}^{-i   m s}\,\psi(x^{0},x^{1},x^{2},x^{3})\,,
\ee
we immediately obtain that the D'Alembert equation for $\varphi$ reduces to  the Schr\"{o}digner equation for $\psi$ given by
\be
2im\partial_{t}\psi\,=\,-\Delta\,\psi\,.
\ee
Analogously to the previous case, we stress that a subgroup of $ISO(1,4)$ preserving the set of functions selected in  Eq.\eqref{eqn: SE constraint} can be identified with the Galilei group.

In other words, we are observing that the D'Alembert equation in the five-dimensional   space-time $(\mathcal{M},\eta)$ is a sort of covering PDE for both the Klein-Gordon and the Schr\"{o}dinger equations in four dimensions. 
Similar considerations were arrived at by other people, starting, however, from completely different motivations \cite{B-G-S-1987,D-B-K-P-1984,G-G-U-2008,O-K-T-O-1989,L-M-S-V-1994}.

Let us now look closer at the geometrical interpretation of the reduction procedure outlined above. 
Our departure point is the five-dimensional Lorentzian manifold $\left( \mathcal{M}, \eta \right)$. 
This manifold can be also read as the five-dimensional Abelian Lie group $\mathbb{R}^5$, equipped with a translation invariant metric tensor, which is also invariant under the pseudo-orthogonal group.
Concerning the way $\mathcal{M}=\mathbb{R}^{5}$ acts on itself, we can consider two different subgroups: one is the subgroup $\mathbb{R}_{sp}$ of translations  along one fixed space-like direction, say $\mathbf{e}_+$, and one is the subgroup $\mathbb{R}_{l}$  of translations along a light-like direction, say $\mathbf{e}_0$. 
What we are going to show is that the space of functions which have been selected in Eq.\eqref{eqn: KG constraint} and Eq.\eqref{eqn: SE constraint} are related to the reduction of the manifold $\mathcal{M}$ with respect to the action of the two subgroups $\mathbb{R}_{sp}$ and $\mathbb{R}_{l}$, respectively. 

In the case of the Klein-Gordon reduction we fix the space-like vector field $\mathbf{e}_+ = \frac{\partial}{\partial x^4}$. 
It is invariant under the action of the Abelian group $\mathcal{M}=\mathbb{R}^5$ on itself, and it generates an action $R^{sp}_g\,\colon\,\mathcal{M}\,\rightarrow\,\mathcal{M}$ of a one-dimensional Abelian subgroup $\mathbb{R}_{sp}\subset\mathbb{R}^{5}=\mathcal{M}$ on $\mathcal{M}$.
Setting $\mathcal{M}_{KG} = \mathcal{M} /\mathbb{R}_{sp}=\mathbb{R}^5 / \mathbb{R}_{sp} \cong \mathbb{R}^4$, the canonical projection $\pi_{KG}\,\colon\,\mathbb{R}^5\,\rightarrow\,\mathcal{M}_{KG}$ defines a principal bundle with structure group $\mathbb{R}_{sp}$. 
Since we are dealing with complex-valued functions on $\mathcal{M}$, we can read the space of maps selected by the condition in Eq.\eqref{eqn: KG constraint} as the space of equivariant complex-valued functions on $\mathcal{M}$. 
If $R^{sp}_{\cdot}(\cdot)\,\colon\,\mathbb{R}_{sp}\times\mathcal{M} \,\rightarrow\,\mathcal{M}$ denotes the action of $\mathbb{R}_{sp}$ on $\mathcal{M}$ generated by $\mathbf{e}_+$, then equivariant functions satisfy the condition
\begin{equation}
\varphi(R_g^{sp}(m)) = \rho(g)^{-1}\varphi(m)
\end{equation}     
where $g\in \mathbb{R}_{sp}$ and $\rho\,\colon\,\mathbb{R}_{sp}\,\rightarrow\,\mathbb{C}$ is a representation of the group $\mathbb{R}_{sp}$ on the vector space $\mathbb{C}$. 
In the situation described by Eq.\eqref{eqn: KG constraint}, the representation  $\rho$ is given by $\rho(a) = \mathrm{e}^{ma}$ with $m\in \mathbb{R}$. 

The space of equivariant complex-valued functions on $\mathcal{M}$ is a module $\Gamma_{KG}(\mathcal{M})$ over the ring $C^{\infty}_{sp}(\mathcal{M})$ of smooth real valued functions which are invariant with respect to the action $R^{sp}_{g}$ of the group $\mathbb{R}_{sp}$ on $\mathcal{M}$. 
This module is isomorphic to the module of sections of the associated vector bundle $F = \left[ \mathcal{M}\times_{\mathbb{R}_{sp}}\mathbb{C} \right]\cong \mathcal{M}_{KG}\times \mathbb{C}$, where $\mathbb{R}_{sp}$ acts on $\mathbb{C}$ via   $\rho$.
Therefore, by imposing the condition in Eq.\eqref{eqn: KG constraint}, we are reducing the initial D'Alembert operator on $\mathcal{M}$ to an operator acting on the sections of the associated vector bundle $F$ with base $\mathcal{M}_{KG}$. 

An analogous procedure can be performed to obtain the Schr\"{o}dinger equation. 
However, in this case we prefer to perform a slightly different reduction procedure in order to preserve the square-integrability of the solutions on the reduced configuration space, which is related to   the physical interpretation of the solutions.
Specifically, we will proceed in such a way to obtain a reduced configuration space, which is the analogue of the bundle $\pi_{KG}\,\colon\,\mathcal{M}\,\rightarrow\,\mathcal{M}_{KG}$, having compact fibers.

The departure point is again the Lorentzian manifold $(\mathcal{M}, \eta)$. 
As before, we consider this manifold as the Abelian Lie group $\mathcal{M}=\mathbb{R}^5$, and we choose a light-like left invariant vector field $\mathbf{e}_0 = \frac{\partial}{\partial x^0} + \frac{\partial }{\partial x^4}$. 
This vector field is the generator of an action of the subgroup $\mathbb{R}_l$ on $\mathcal{M}$. 
Let us consider the discrete subgroup  $\mathbb{Z}_l\subset \mathbb{R}_l$ of translation of a given quantity $2\pi$. 
The first step of the procedure consists in going from the manifold $\mathcal{M}$ to the manifold $\tilde{\mathcal{M}} := \mathcal{M} / \mathcal{Z}_l \cong U(1)\times \mathbb{R}^4$. 
Since the metric tensor is invariant with respect to $\mathbb{Z}_l$, there exist a metric tensor $\tilde{\eta}$ on the quotient manifold $\tilde{\mathcal{M}}$ such that $\eta = \tilde{\pi}^*(\tilde{\eta})$. 
Here $\tilde{\pi}\,\colon\,\mathcal{M}\,\rightarrow\,\tilde{\mathcal{M}}$ denotes the projection from $\mathcal{M}$ to the quotient manifold $\tilde{\mathcal{M}}$.
The space of functions $C^{\infty}(\tilde{\mathcal{M}})$ can be identified with the subspace $C^{\infty}_p(\mathcal{M})$ of functions on $\mathcal{M}$ which are invariant with respect to $\mathbb{Z}_l$, i.e., they are periodic of period $2\pi$ along the integral curves of the vector field $\mathbf{e}_0$. 
Since the derivative of a periodic function is a periodic function, the vector field $\mathbf{e}_0$ is projectable and there is a vector field $\tilde{\mathbf{e}}_0$ on $\tilde{\mathcal{M}}$ which is $\tilde{\pi}$-related to $\mathbf{e}_0$. 
This vector field is the generator of an action, say $R^l_g\,\colon\,\tilde{\mathcal{M}}\,\rightarrow\,\tilde{\mathcal{M}}$, of the group $U(1)$ on $\tilde{\mathcal{M}}$. 
Now, we can complete the reduction procedure defining the reduced configuration space $\mathcal{M}_S := \tilde{\mathcal{M}} / U(1)$ and the projection $\pi_S\,\colon\,\tilde{\mathcal{M}} \,\rightarrow\,\mathcal{M}_S$. 
As mentioned before, this projection defines a bundle structure with base manifold $\mathcal{M}_{S}\cong \mathbb{R}^4$ and compact fibers diffeomorphic to the Lie group $U(1)$. 
The complex-valued functions which satisfy the condition in Eq.\eqref{eqn: SE constraint} are invariant under the action of the group $\mathbb{Z}_l$, and we can think of them as the pullback to $\mathcal{M}$ of functions on $\tilde{\mathcal{M}}$. 
Furthermore,   they are equivariant with respect to the action $R^l_{\cdot}(\cdot)\,\colon\,U(1)\times\tilde{\mathcal{M}}\,\rightarrow \,\tilde{\mathcal{M}}$ of the group $U(1)$ generated by the vector field $\mathbf{e}_0$. 
In this case, the representation $\rho\,\colon\,U(1)\,\rightarrow\,\mathbb{C}$ of the group $U(1)$ is given by $\rho(s_0) = \mathrm{e}^{im s}$.
These functions form a module $\Gamma_S(\tilde{\mathcal{M}})$ with respect to the ring $C^{\infty}_l(\tilde{\mathcal{M}})$ of functions which are invariant with respect to the action $R^l_{g}$. 
This module is isomorphic to the module of sections of the associated vector bundle $F = [\tilde{\mathcal{M}}\times_{U(1)}\mathbb{C}]\cong \mathcal{M}_{S}\times\mathbb{C}$ and the D'Alembert operator on $\tilde{\mathcal{M}}$ reduces to the Schr\"{o}dinger operator when acting on the module $\Gamma_S(\tilde{\mathcal{M}})$.

\section{Principal symbol and particle dynamics}\label{sec: Principal symbol and particles}

The principal symbol  associated with the differential operator defining the D'Alembert equation coincides with the contravariant counterpart of the metric tensor $\eta$.
Accordingly, we may look at $\sigma_{DA}$ as defining a function on $T^{*}\mathcal{M}$ when contracted with the tautological 1-form $\theta_{0}$ of $T^{*}\mathcal{M}$ twice.
Specifically, we obtain
\be
\sigma_{DA}\,=\,\eta(\theta_{0},\theta_{0})\,=\,\eta^{\mu\nu}p_{\mu}p_{\nu}\,=\,2p_{t}p_{s} - \delta^{jk}p_{j}p_{k}\,.
\ee
Setting $\sigma_{DA}=0$, we obtain a sort of ``mass-shell'' in 5 dimensions.
Without the origin, we call this submanifold $\Sigma_{DA}$.
Now, both the vector field $\mathbf{e}_{+}$ and $\mathbf{e}_0$ may be used to perform a reduction on $\Sigma_{DA}$ that will lead to the relativistic mass-shell in the former case, and to the non-relativistic mass-shell in the second case.

Specifically, let us start from the cotangent bundle $T^*\mathcal{M}$ of the configuration space $\mathcal{M}$. 
It is a symplectic manifold equipped with the canonical symplectic form $\omega = \mathrm{d} \theta_0$. 
Let us consider the action on $\mathcal{M}$ generated by the vector field $\mathbf{e}_+$. 
The cotangent lift $\mathbf{e}_+^{\uparrow}$ of the vector field $\mathbf{e}_+$ generates a flow on $T^*\mathcal{M}$ which preserves both $\omega$ and $\theta_0$. 
There is a momentum map $J_+\,\colon\,T^*\mathcal{M}\,\rightarrow\,\mathbb{R}$ associated with this action given by
\begin{equation}
J_+ = i_{\mathbf{e}^{\uparrow}_+}\theta = p_4\,.
\end{equation}
Choosing the regular value $m$ for this function, by the Marsden-Weinstein reduction theorem \cite{AbraMars-Foundations_of_Mechanics}, the inverse image $J_+^{-1}(m)$ is a submanifold of $T^*\mathcal{M}$ and the vector field $\mathbf{e}^{\uparrow}_+$ is still tangent to it. 
Let $\Sigma_+\subset T^*\mathcal{M}$ be the submanifold of $\mathcal{M}$ defined by the conditions 
\begin{equation}\label{mass-shell_reduction_KG}
J_+=m \,,\quad \sigma_{DA}=0\,.
\end{equation}
Since $\mathbf{e}^{\uparrow}_+$ is tangent to $\Sigma_+$, we can build the reduced mass-shell $\Sigma_m = \Sigma_+ / \mathbb{R}_{sp}$. 
It is straightforward to see that this space is diffeomorphic to the mass-shell of a relativistic particle in $T^*\mathcal{M}_{KG}$, since Eqs.\eqref{mass-shell_reduction_KG} can be rewritten as 
\begin{equation}
p_0^2-\sum_{j=1}^3 p_j^2 = m^2\,.
\end{equation}

Similarly, the cotangent lift $\mathbf{e}_0^{\uparrow}$ of the vector field $\mathbf{e}_0$ acts on $T^*\mathcal{M}$ preserving $\omega$.
Let us now introduce the retarded and advanced coordinates on $\mathcal{M}$ and let $\left\lbrace p_s = \frac{1}{2}\left( p_0 + p_4 \right), p_t = \frac{1}{2}\left( p_0-p_4 \right), p_1,p_2,p_3 \right\rbrace$ be the associated cotangent bundle coordinates. 
In this case, the momentum map associated with the action generated by $\mathbf{e}_0^{\uparrow}$ is given by the function $J_0\,\colon\,T^*\mathcal{M}\,\rightarrow\,\mathbb{R}$ 
\begin{equation}
J_0 = p_s
\end{equation}
As before, we can select the submanifold $\Sigma_0 \subset T^*\mathcal{M}$ which is defined by the following conditions
\begin{equation}\label{mass-shell_reduction_S}
J_0 = m \,\quad \sigma_{DA} = 2p_sp_t -\sum_{j=1}^3 p_j^2 = 0\,.
\end{equation}
Then, the vector field $\mathbf{e}_0^{\uparrow}$ is tangent to $\Sigma_0$ and we obtain the reduced mass-shell $\Sigma_E$. 
However, differently from the previous case, the manifold $\Sigma_E$ is diffeomorphic to the submanifold of $T^*\mathcal{M}_{S}$ defined by the non-relativistic ``mass-shell'' relation, since the two relations in Eq.\eqref{mass-shell_reduction_S} can be rewritten as
\begin{equation}
2m p_t - \sum_{j=1}^3 p_j^2 = 0\,.
\end{equation} 
It is immediate to check that $p_t$ defines the kinetic energy of a particle in a three-dimensional Euclidean space.

\section{Reduction in the variational formulation}\label{eqn: Reduction in the variational formulation}

All the differential equations we have considered so far admit of a variational formulation in terms of an action principle.
The variational formulation is very convenient because it allows for the definition of Poisson Brackets (Peierls Brackets) \cite{A-C-DC-I-2017,A-C-DC-I-M-2017,C-DC-I-M-S-2020-01,C-DC-I-M-S-2020-02}
 and is a preliminary step for quantization.
Using the variational formulation, we shall now provide  a description of the reduction mechanism for the Schr\"{o}dinger and (complex) Klein-Gordon equations in terms of the multisymplectic formulation of the Schwinger-Weiss action principle.

Let us start with the five-dimensional, globally hyperbolic space-time $(\mathcal{M}, \eta)$  introduced in section \ref{sec: reduction of e.o.m.}.
Let $\pi_0 \colon E \rightarrow \mathcal{M}$ be a Hermitean vector bundle over $\mathcal{M}$, with $\pi_{0}$ the standard projection on the second factor.   
The standard fibre $V$ of the bundle will be a finite-dimensional complex linear space that, for the purposes of this contribution will be just one-dimensional.
Since $\mathcal{M}$ is contractible, the bundle is trivalizable, that is, we have $E\cong\mathbb{C}\times \mathcal{M}$.
If $(x^{\mu})$ is a (global) set of Cartesian coordinates on $\mathcal{M}$, then $(x^\mu, u, \bar{u})$, $u \in \mathbb{C}$, will define natural adapted coordinates for the bundle $E$. 
The fields are sections $\phi$ of the bundle $\pi_0 \colon E \rightarrow  \mathcal{M}$, and since   $E$ is trivializable, the sections can be identified with functions $\phi \colon \mathcal{M} \to \mathbb{C}$.  


The  covariant phase space $\mathcal{P}$  is a fibre bundle over both $E$ and the spacetime $\mathcal{M}$ \cite{CarinCrampIbort1991-Multisymplectic,IbSpiv2017-Covariant_Hamiltonian_boundary}.
Technically speaking, $\mathcal{P}$ is the (affine) dual bundle of the vertical bundle $V(E)\subset \mathbf{T}E$, and, since the bundle $E$ describing the  theory is trivializable, $\mathcal{P}$  is diffeomorphic to $\mathbf{T}^\mathbb{C}\mathcal{M}\times \mathbb{C}$.   
Adapted coordinates for the covariant phase space are given by $(x^\mu, u , \bar{u}; \rho^\mu, \bar{\rho}^\mu)$, and the projection $\pi_{E}\,:\, \mathcal{P}\,\rightarrow \, E$ is given  by $\pi_{E}(\rho^{\mu}, \bar{\rho}^\mu,u, \bar{u},x^{\mu})  = (u, \bar{u},x^{\mu})$, while the projection $\pi\colon \mathcal{P}\,\rightarrow \, \mathcal{M}$ is given by $\pi(\rho^{\mu}, \bar{\rho}^\mu,u, \bar{u},x^\mu) = (x^{\mu})$.

Let $\mathscr{F}_{\mathcal{P}}$ denote the space of sections $\chi\,:\, \mathcal{M}\,\rightarrow\,\mathcal{P}$ of the projection $\pi$, such that the compostion $\pi_E \circ \chi = \phi$ is a section of $\pi_0$. 
We write $\chi$ as $(\phi, P)$ and   
\begin{equation}
\chi(x^{\mu}) = (x^{\mu}, \phi(x^{\mu}), \bar{\phi}(x^\mu), P^{\nu}(x^{\mu}), \bar{P}^{\nu}(x^{\mu}))\,, \quad \mu,\nu = 0,1,\ldots,m\,.
\end{equation}
Elements $\chi = (\phi, P)$ in $\mathscr{F}_{\mathcal{P}}$ will be the fields of the theory.
The analytical background of the theory can be established by assuming specific regularity conditions for the fields $\chi$ \cite{C-DC-I-M-S-2020-02}.   

A variation $\delta \chi$ for $\chi\in\mathscr{F}_{\mathcal{P}}$ is a tangent vector to $\mathscr{F}_{\mathcal{P}}$ at  $\chi$, which may be identified with a vector field $U_{\chi}$ along $\chi$ on $\mathcal{P}$, i.e., $U_{\chi} (\chi(x)) = \frac{d}{ds}\mid_{s = 0} \chi_s(x)$,
where $\chi_s \colon \mathcal{M} \to \mathcal{P}$, $-\epsilon < s <\epsilon$, is a smooth curve of fields such that $\chi_0 = \chi$. Clearly,
$U_{\chi}$ is vertical with respect to the fibration $\pi\colon\mathcal{P}\ra\mathcal{M}$.
Thus, the tangent space $\mathbf{T}_{\chi}\mathscr{F}_{\mathcal{P}}$ is given by all such $\delta \chi = U_{\chi}$.
In the following, it will be useful to extend $U_{\chi}$ to a vertical vector field $\widetilde{U}$ in a neighbourhood of the image of $\chi$ inside $\mathcal{P}$ that will be given by (note that in spite of $E$ being a complex line bundle, the covariant phase space $\mathcal{P}$ is a real manifold and, consequently, vector fields on it must be real)
\begin{equation}
\tilde{U} = \bar{U}_{\phi} \frac{\partial }{\partial u} + U_{\phi} \frac{\partial }{\partial \bar{u}}  + \bar{U}_{P}^{\mu}\frac{\partial }{\partial \rho^{\mu}} + U_{P}^{\mu}\frac{\partial }{\partial \bar{\rho}^{\mu}}\,.
\end{equation}

The dynamics of the theory will be described by using the Schwinger-Weiss action principle \cite{C-DC-I-M-S-2020-01,C-DC-I-M-S-2020-02}.
Given a Hamiltonian function $H\colon \mathcal{P}\,\rightarrow\,\mathbb{R}$ and volume form $\mathrm{vol}_\mathcal{M} = \dd^{5}x = \dd x^0\wedge \dd x^1 \wedge \cdots \wedge \dd x^{4}$, on the base manifold $\mathcal{M}$, we consider the  $m+1$-form\footnote{It is possible to define this form in an intrinsic way, see for instance \cite{IbSpiv2017-Covariant_Hamiltonian_boundary}.} $\theta_H$  on $\mathcal{P}$ given by
\begin{equation}\label{eq:thetaH}
\theta_H = (\bar{\rho}^{\mu}\dd u + \rho^\mu \dd \bar{u}) \wedge i_{\frac{\partial}{\partial x^{\mu}}}\mathrm{vol}_\mathcal{M} - H \mathrm{vol}_\mathcal{M}\,.
\end{equation}
The Hamiltonian function will have the form
\begin{equation}\label{eq:hamiltonian}
H =   \eta_{\mu \nu} \bar{\rho}^{\mu}\rho^{\nu} . 
\end{equation}
The action functional $S\colon\mathscr{F}_{\mathcal{P}}\, \rightarrow\, \mathbb{R}$, of the theory can be written as
\begin{equation}
S[\chi] = \int_{\mathcal{M}}\chi^*\left( \theta_H \right) = \int_{\mathcal{M}}\left( \bar{P}^{\mu}\partial_\mu \phi + P^{\mu}\partial_\mu \bar{\phi} - H \right) \mathrm{vol}_\mathcal{M}\,.
\label{action_klein-gordon}
\end{equation}
Following what is done in Ref.\cite{C-DC-I-M-S-2020-01}, given  $U_{\chi}\in\mathbf{T}_{\chi}\mathscr{F}_{\mathcal{P}}$, the variation  $\mathrm{d}S[\chi](U_{\chi})$ of $S$ is 
\be
\begin{split}
\mathrm{d}S[\chi](U_{\chi})&=  \mathbb{EL}_{\chi}(U_{\chi}) + \int_{\partial \mathcal{M} } \chi_{\partial\mathcal{M}}^*\left( i_{\tilde{U}}\theta_H \right),
\end{split}
\ee
where $\tilde{U}$ is any extension of $U_{\chi}$, and
\begin{equation} 
\mathbb{EL}_{\chi}(U_{\chi}) := \int_{\mathcal{M} }\chi^*\left( i_{\tilde{U}}\dd \theta_H \right) .
\end{equation}
The Schwinger-Weiss action principle states that the variations of the action depend solely on the variations of the fields at the boundary, hence, being $\partial\mathcal{M}=\emptyset$,  the actual dynamical configurations of the fields of the theory must satisfy the Euler-Lagrange equations
$$
\frac{\partial \phi}{\partial x^\mu} = \frac{\partial H}{\partial \bar{P}^\mu} \, ,\qquad  \frac{\partial \bar{\phi}}{\partial x^\mu} = \frac{\partial H}{\partial P^\mu} \, , \qquad   \frac{\partial P^\mu}{\partial x^\mu} = - \frac{\partial H}{\partial \bar{\phi}} \, , \qquad   \frac{\partial \bar{P}^\mu}{\partial x^\mu} = - \frac{\partial H}{\partial \phi} \, ,
$$
which can be geometrically interpreted as the zeroes of a 1-form $\mathbb{EL}$ on the space of fields $\mathscr{F}_{\mathcal{P}}$ given by
\begin{equation}\label{eqn: Schwinger-Weiss action principle 1}
\mathbb{EL}_{\chi}(U_{\chi}) := \int_{\mathcal{M} }\chi^*\left( i_{\tilde{U}}\dd \theta_H \right) = 0 \,,\quad \forall U_{\chi}\in\mathbf{T}_{\chi}\mathscr{F}_{\mathcal{P}}\, .
\end{equation}
We note that such a geometrical reformulation of the Schwinger-Weiss variational principle should be used to introduce a Quantum Action Principle in the groupoid reformulation of Schwinger's algebra of selective measurements which has been recently proposed by some of the authors (see \cite{C-I-M-2018,C-I-M-02-2019,C-I-M-03-2019,C-I-M-05-2019,C-DC-I-M-2020,C-DC-I-M-02-2020} for more details).

We will denote by $\mathcal{EL}_{\mathcal{M}}\subset \mathscr{F}_{\mathcal{P}}$ the space of solutions of Euler-Lagrange equations
\begin{equation}
\mathcal{EL}_{\mathcal{M}}\,:=\,\left\{\chi\in \mathscr{F}_{\mathcal{P}}\,\colon\,\mathbb{EL}_{\chi}(U_\chi) = 0\,,\quad \forall U_{\chi} \in \mathbf{T}_{\chi}\mathscr{F}_{\mathcal{P}}\right\}.
\end{equation} 
From (\ref{eq:hamiltonian}), we obtain the Euler-Lagrange equations  (also known as the de Donder-Weyl equations)   
\begin{equation}
\frac{\partial \phi}{\partial x^{\mu}} = \eta_{\mu \nu} P^{\nu} \, , \qquad
\frac{\partial P^{\mu}}{\partial x^{\mu}} = 0 \,,
\end{equation}
and their complex conjugate.

\subsection{Klein-Gordon equation}

The reduction procedure will consist of selecting a subspace of the space of field $\mathscr{F}_\mathcal{P}$. 
Analogously to what is done in section \eqref{sec: reduction of e.o.m.}, we select the space-like vector field $\frac{\partial}{\partial x^{4}}$, and  we first consider the subspace of sections of $E$ the elements of which can be written as
\be\label{eq:homogeneous KG}
\begin{split}
\phi (x^{0}, x^1,x^{2}, x^{3},x^{4}) &= e^{-mx^{4}} \,\psi (x^{0}, x^1, x^{2}, x^{3} ) \\
\overline{\phi} (x^{0}, x^1,x^{2}, x^{3},x^{4}) &= e^{-m x^{4}}\, \overline{\psi} (x^{0}, x^1, x^{2}, x^{3} ) \, ,
\end{split}
\ee
with $m > 0$ a constant.
The form of the fields $\phi$ determined by Eq. (\ref{eq:homogeneous KG}) affects the form of the associated momenta fields.
Recalling that the coordinates $\rho^\mu$ are dual to the coordinates of 1-jets $j_x^1 \phi = (x, \phi (x), \partial_\mu \phi)$, being $\partial_0 \phi = e^{-m x^{4}} \partial_0 \psi$, $\partial_k \phi = e^{-m x^{4}} \partial_k \psi$, and $\partial_4 \phi = -m e^{-m x^{4}}\psi$, we get that
\be
\begin{split}
P^0 (x^{0},x^k,x^{4}) & = e^{- m x^{4}} \pi^0 (x^{0},x^k)   \\
P^k (x^{0},x^k,x^{4}) &= e^{-m x^{4}} \pi^k(x^{0},x^k)  \\
P^4 (x^{0},x^k,x^{4}) &= -m e^{-m x^{4}} \lambda(x^{0},x^k) \, ,
\end{split}
\ee
and their complex conjugate.
The space of all such fields is denoted by $\mathcal{F}_{KG}$, and, for the sake of notational simplicity, a generic element in $\mathcal{F}_{KG}$ is denoted by $\chi=(\psi,\pi_{0},\pi_{k},\lambda)$ omitting the dependence on $x^{4}$ and on the complex conjugate fields.
Then, the restriction of the action $S$ to the space of fields $\mathcal{F}_{KG}$ becomes
\begin{eqnarray*}
S  (\psi, \pi_{0},\pi_{k} , \lambda ) &=&  \int_\mathcal{M} e^{-2m x^{4}}\left( \bar{\pi}^0 \partial_0 \psi +  \pi^0 \partial_0 \bar{\psi} +  m^{2} \bar{\lambda}\psi  + m^{2}  \lambda \bar{\psi}    \right. \\ &&  + \left. \bar{\pi}^k \partial_k \psi + \pi^k \partial_k \bar{\psi}  - \bar{\pi}^{0}\pi^{0}  + \delta_{ij} \bar{\pi}^i \pi^j  + m^{2}|\lambda|^{2}  \right) \mathrm{vol}_\mathcal{M} \, .
\end{eqnarray*}
The convergence of the previous integral may be handled by specifying a suitable regularity conditions for the fields, or by a careful ``renormalized'' definition of the action.
As before, the dynamics may be described in terms of the Schwinger-Weiss action principle for sections $\chi = (\psi, \pi_{0},\pi_{k}, \lambda)$. 
The result is the following system of de Donder-Weyl equations
\be
\begin{split}
\lambda +  \psi &=0\, , \quad \bar{\lambda} +    \bar{\psi}=0 , \\
\partial_0  \psi   - \pi^{0} & =0 \,,\quad \partial_0 \bar{\psi}    - \bar{\pi}^{0}=0 ,\\
\partial_k  \psi  + \delta_{ij}  \pi^i &=0 \,, \quad \partial_k \bar{\psi} + \delta_{ij} \bar{\pi}^i =0 ,\\
\partial_{0}\bar{\pi}_{0} - m^{2}\bar{\lambda} + \partial_{k}\bar{\pi}_{k} &=0 \,,\quad \partial_{0} \pi_{0} - m^{2} \lambda + \partial_{k} \pi_{k}=0 \,.
\end{split}
\ee
A direct computation shows that the previous system of equations reduces to the (complex) Klein-Gordon  equation for $\psi$ and its complex conjugate given by
\be
\partial_{0}^{2}\psi - \,\Delta\psi + m^{2}\psi\,=\,0 \,,\quad  \partial_{0}^{2}\bar{\psi} - \Delta\psi + m^{2}\bar{\psi}\,=\,0\,.
\ee

\subsection{Schr\"{o}dinger equation}

In this subsection,  it will be shown that the Schr\"odinger equation (SE) can be obtained in a rather natural way as a reduction of the complex D'Alembert field introduced above. 
In this sense, it will be shown that the SE can be interpreted not as a non-relativistic limit, but rather as a genuine reduction of the D'Alembert fields in a larger dimensional space-time.   
The presentation will be necessarily sketchy and further details and implications of the presented results will be discussed elsewhere.   

As done in section \ref{sec: reduction of e.o.m.}, in the globally hyperbolic spacetime $(\mathcal{M}, \eta)$, we consider the  ``advanced'' and ``retarded coordinates''  $t,s$ give in Eq.\eqref{eqn: coordinate cono luce}, and the resulting expression of $\eta$ given in Eq.\eqref{eqn: metrica nelle coordinate di cono luce}.
We now consider  the multisyplectic formulation of D'Alembert theory on $\mathcal{M}$ given above in the new coordinates $t,s,x^k$.
The Hamiltonian $H$ given by Eq. (\ref{eq:hamiltonian})   becomes 
\be
H =  \bar{\rho}^t \rho^s + \rho^t \bar{\rho}^s - \eta_{ij} \bar{\rho}^i \rho^j   \, .
\ee
The corresponding action of the theory on the space of fields $\mathcal{F}_\mathcal{P}$ will have the expression:
\be
S (\chi ) = \int_\mathcal{M} \bar{P}^t \partial_t \phi +  P^t \partial_t \bar{\phi}  + \bar{P}^s \partial_s \phi +  P^s \partial_s \bar{\phi} - (\bar{P}^t P^s + P^t \bar{P}^s) + \eta_{ij} \bar{P}^i P^j   \, ,
\ee
where $P^t$, $P^s$ represent the components of the momenta fields of the theory in the directions of $t$ and $s$ respectively.  

The reduction procedure will consist of selecting a subspace of the space of field $\mathcal{F}_\mathcal{P}$. 
Analogously to what is done in section \eqref{sec: reduction of e.o.m.}, we select the light-like vector field $\frac{\partial}{\partial s}$, and  we first consider the subspace of sections of $E$ the elements of which can be written as
\be\label{eq:homogeneous}
\begin{split}
\phi (t,s, x^1,x^{2}, x^{3}) &= e^{ims} \,\psi (t, x^1, x^{2}, x^{3} ) \\
\overline{\phi} (t,s, x^1,x^{2}, x^{3}) &= e^{-ims}\, \overline{\psi} (t, x^1, x^{2}, x^{3} ) \, ,
\end{split}
\ee
with $m > 0$ a constant.
The form of the fields $\phi$ determined by Eq. (\ref{eq:homogeneous}) affects the form of the associated momenta fields.
Recalling that the coordinates $\rho^\mu$ are dual to the coordinates of 1-jets $j_x^1 \phi = (x, \phi (x), \partial_\mu \phi)$, being $\partial_t \phi = e^{ism} \partial_t \psi$, $\partial_k \phi = e^{ism} \partial_k \psi$, and $\partial_s \phi = im e^{ism}\psi$, we get that
\be
\begin{split}
P^t (t,s,x^k) & = e^{ism} \pi^t (t,x^k)   \\
P^k (t,s,x^k) &= e^{ism} \pi^k(t,x^k)  \\
P^s (t,s,x^k) &= im e^{ism} \lambda(t,x^k) \, ,
\end{split}
\ee
and their complex conjugate.
The space of all such fields is denoted by $\mathcal{F}_{SE}$, and, for the sake of notational simplicity, a generic element in $\mathcal{F}_{SE}$ is denoted by $\chi=(\psi,\pi_{t},\pi_{k},\lambda)$ omitting the dependence on $s$ and on the complex conjugate fields.

Then, the restriction of the action $S$ to the space of fields $\mathcal{F}_{SE}$ becomes
\begin{eqnarray*}
S  (\psi, \pi_{t},\pi_{k} , \lambda ) &=&  \int_\mathcal{M} \left( \bar{\pi}^t \partial_t \psi +  \pi^t \partial_t \bar{\psi}  - i m \bar{\lambda} (i m \psi + \pi^t) + i m  \lambda ( -i m \bar{\psi}  + \bar{\pi}^t ) \right. \\ && + \left. \bar{\pi}^k \partial_k \psi + \pi^k \partial_k \bar{\psi}  + \delta_{ij} \bar{\pi}^i \pi^j    \right) \mathrm{vol}_\mathcal{M} \, .
\end{eqnarray*}
Analogously to the Klein-Gordon case, the convergence of the previous integral may be handled by specifying a suitable regularity conditions for the fields, or by a careful ``renormalized'' definition of the action.
As before, the dynamics may be described in terms of the Schwinger-Weiss action principle for sections $\chi = (\psi, \pi_{t},\pi_{k}, \lambda)$. 
The result is the following system of de Donder-Weyl equations
\be
\begin{split}
\pi^t + i m \psi &=0\, , \quad \bar{\pi}^t -  i m \bar{\psi}=0 , \\
\partial_t  \psi   + i m  \lambda & =0 \,,\quad \partial_t \bar{\psi}  - i m \bar{\lambda}=0 ,\\
\partial_k  \psi  + \delta_{ij}  \pi^i &=0 \,, \quad \partial_k \bar{\psi} + \delta_{ij} \bar{\pi}^i =0 ,\\
\partial_{t}\bar{\pi}_{t} - m^{2}\bar{\lambda} + \partial_{k}\bar{\pi}_{k} &=0 \,,\quad \partial_{t} \pi_{t} - m^{2} \lambda + \partial_{k} \pi_{k}=0 \,.
\end{split}
\ee
A direct computation shows that the previous system of equations reduces to the Schr\"{o}dinger equation for $\psi$ and its complex conjugate given by
\be
2i m\partial_{t}\psi\,=\,- \,\Delta\psi \,,\quad 2i m \partial_{t}\bar{\psi}\,=\, \Delta\bar{\psi}\,.
\ee

\section{Conclusions}

We presented different instances of reduction procedures providing the Schr\"{o}dinger and Klein-Gordon equations.
Specifically, we presented a reduction procedure at the level of equations of motion in section   \ref{sec: reduction of e.o.m.}, and a reduction procedure at the level of the variational description encoded in the Schwinger-Weiss action principle in the multisymplectic formalism in section \ref{eqn: Reduction in the variational formulation}.
In all these cases, the reduction   starts from the D'Alembert equation on a five-dimensional flat Lorentzian spacetime, and arrives at the Schr\"{o}dinger equation by means of a suitable reduction associated with a light-like vector field, and  at the Klein-Gordon equation by means of a suitable reduction associated with a space-like vector field.

Motivated by these instances, it is reasonable to try to understand if  this reduction picture may be extended to a more general setting.
Indeed, we will now argue that a similar procedure is indeed possible for the Dirac equation, at least at the level of the reduction of the equations of motion as presented in section \ref{sec: reduction of e.o.m.}.
However, we will have to slighly change our approach and consider an  unfolding space which has double the dimensions of the spacetime on which the Dirac equation is defined.
Specifically,  there is a reduction procedure that takes us from the (complex) ``Klein-Gordon'' equation on a suitable 8-dimensional spacetime to the Dirac equation on Minkowski spacetime.
At this purpose, let us consider the standard Minkowski spacetime $(\mathbb{M}^{1,3}, \eta)$, where $\mathbb{M}^{1,3}=\mathbb{R}^{4}$, and $\eta$ is  the metric tensor given by  $\eta =  - \dd y^0 \otimes \dd y^0 + \delta_{jk}\dd y^j \otimes \dd y^k$, with respect to the global set of Cartesian coordinates $(y^{\mu})$, $\mu = 0,1,2,3$.  
Then, consider the ``anti-Minkowski spacetime'' $(\overline{\mathbb{M}}^{1,3}, \overline{\eta})$, where $\overline{\mathbb{M}}^{1,3}=\mathbb{R}^{4}$, and $\overline{\eta}$ is  the metric tensor given by  $\overline{\eta} =    \dd \overline{y}^0 \otimes \dd \overline{y}^0 -\delta_{jk}\dd \overline{y}^j \otimes \dd \overline{y}^k$, with respect to the global set of Cartesian coordinates $(\overline{y}^{\mu})$, $\mu = 0,1,2,3$.  
Now, we form the $8$-dimensional ``spacetime'' $(\mathcal{M},g)$, where $\mathcal{M}=\mathbb{M}^{1,3}\times\overline{\mathbb{M}}^{1,3}$, and $g=\eta\oplus\overline{\eta}$, that is, $g= \dd y^0 \otimes \dd y^0 - \delta_{jk}\dd y^j \otimes \dd y^k  -\dd \overline{y}^0 \otimes \dd \overline{y}^0 + \delta_{jk}\dd \overline{y}^j \otimes \dd \overline{y}^k$.
Let us introduce the ``advandeced'' and ``retarded'' coordinates
\be
\begin{split}
x^0=y^{0} - \overline{y}^{0},\quad x^{1}=y^{1} - \overline{y}^{1},\quad x^{2}=y^{2} - \overline{y}^{2},\quad x^{3}=y^{3} - \overline{y}^{3},  \\
s=y^{0} + \overline{y}^{0},\quad \xi^{1}=y^{1} + \overline{y}^{1},\quad \xi^{2}=y^{2} + \overline{y}^{2},\quad \xi^{3}=y^{3} + \overline{y}^{3} .
\end{split}
\ee
With respect to these coordinates, the metric tensor $g$ reads
\be
g\,=\,-\dd x^0\otimes_{S}\dd s + \dd x^{1}\otimes_{S}\dd \xi^{1} +\dd x^{2}\otimes_{S}\dd \xi^{2} +\dd x^{3}\otimes_{S}\dd \xi^{3}\,.
\ee
Now, we consider the complex Klein-Gordon equation for  a $\mathbb{C}^{4}$-valued function $\Phi$ as determined by the metric tensor $g$.
Specifically, we consider the equation
\be\label{eqn: 8 KG}
\left(-\frac{\partial}{\partial x^0} \frac{\partial }{\partial s} + \frac{\partial}{\partial x^{1}} \frac{\partial }{\partial \xi^{1}} +\frac{\partial}{\partial x^{2}} \frac{\partial }{\partial \xi^{2}} +\frac{\partial}{\partial x^{3}} \frac{\partial }{\partial \xi^{3}}\right)\Phi + m^{2}\Phi\,=\,0.
\ee
Then, in analogy with what we did in section \ref{sec: reduction of e.o.m.}, we focus on the subspace of functions $\Phi$ given by
\be\label{eqn: constraint Dirac}
\Phi(x^0,x^{1},x^{2},x^{3},s,\xi^{1},\xi^{2},\xi^{3})\,=\,e^{i m(s\gamma_{0} -\xi^{1}\gamma_{1} - \xi^{2}\gamma_{2} - \xi^{3}\gamma_{3})}\cdot\Psi(x^0,x^{1},x^{2},x^{3}),
\ee
where $\Psi$ is a $\mathbb{C}^{4}$-valued function, and the $\gamma$'s are the $(4\times 4)$ Dirac matrices characterized by
\be
\gamma_{\mu}\gamma_{\nu} + \gamma_{\nu}\gamma_{\mu}\,=\,\eta_{\mu\nu}.
\ee
At this point, we can substitute Eq.\eqref{eqn: constraint Dirac} into   Eq.\eqref{eqn: 8 KG}, and a direct computation leads to 
\be
e^{i m(s\gamma_{0} -\xi^{1}\gamma_{1} - \xi^{2}\gamma_{2} - \xi^{3}\gamma_{3})}\,\cdot\,\left[-i  (\gamma_{0}\partial_{0} - \gamma_{1}\partial_{1} - \gamma_{2}\partial_{2} - \gamma_{3}\partial_{3})\Psi + m \Psi\right]\,=\,0 \,,
\ee
which is clearly equivalent to the Dirac equation for $\Psi$
\be
\left(i  \slashed{\partial} - m\right) \Psi\,=\,0	\,.
\ee
Of course, a more detailed analysis of this reduction is needed, for instance to understand the geometrical aspects behind the ``doubling'' of the spacetime dimensions, to understand a possible reduction at the level of principal symbols (and associated ``particles'') along the lines of what is done in section \ref{sec: Principal symbol and particles}, and to undestand a possible variational formulation in the multisymplectic framework along the lines of what is done in section \ref{eqn: Reduction in the variational formulation}.
A procedure similar to the one giving rise to Schr\"{o}dinger equation should provide us with the Pauli equation for spinning particles. 
The intrinsic formulation of the problem should allow more easily a generalization to non-flat spacetime.
We postpone this analysis to a future publication.

\section*{Acknowledgments}

F.D.C. and A.I. would like to thank partial support provided by the MINECO research project MTM2017-84098-P and QUITEMAD++, S2018/TCS-A4342. A.I. and G.M. acknowledge financial support from the Spanish Ministry of Economy and Competitiveness, through the Severo Ochoa Programme for Centres of Excellence in RD(SEV-2015/0554). G.M. would like to thank the support provided by the Santander/UC3M Excellence Chair Programme 2019/2020, and he is also a member of the Gruppo Nazionale di Fisica Matematica (INDAM), Italy.

%

\end{document}